 \documentclass[final,1p,times,twocolumn]{elsarticle}

\usepackage{graphicx}
\usepackage{amssymb}

\usepackage{lineno}
\usepackage{multirow}
\usepackage{bigstrut}

\usepackage{color}
\usepackage{ulem}
\usepackage{textcomp}

\journal{Spectrochimica Acta  B}

\begin{document}

\begin{frontmatter}


\title{Characterization of sub-monolayer coatings as novel calibration samples for X-ray spectroscopy}



\author[ptb]{Philipp H\"onicke}
\author[axo]{Markus Kr\"amer}
\author[tub]{Lars L{\"u}hl}
\author[ixo]{Konstantin Andrianov}
\author[ptb]{Burkhard  Beckhoff}
\author[axo]{Rainer Dietsch}
\author[axo]{Thomas Holz}
\author[tub]{Birgit Kanngie{\ss}er}
\author[axo]{Danny Wei{\ss}bach} 
\author[ixo]{Thomas Wilhein}
\address[ptb]{Physikalisch-Technische Bundesanstalt (PTB), Abbestr. 2-12, 10587 Berlin, Germany}
\address[axo]{AXO DRESDEN GmbH, Gasanstaltstr. 8b, 01237 Dresden, Germany}
\address[tub]{Technische Universit{\"a}t Berlin, IOAP, Hardenbergstr. 36, 10623 Berlin, Germany}
\address[ixo]{Hochschule Koblenz, RheinAhrCampus, Joseph-Rovan-Allee 2, 53424 Remagen, Germany}

\begin{abstract}
With the advent of both modern X-ray fluorescence (XRF) methods and improved analytical reliability requirements the demand for suitable reference samples has increased. Especially in nanotechnology with the very low areal mass depositions, quantification becomes considerably more difficult. However, the availability of suited reference samples is drastically lower than the demand.
Physical vapor deposition (PVD) techniques have been enhanced significantly in the last decade driven by the need for extremely precise film parameters in multilayer production. We have applied those techniques for the development of layer-like reference samples with mass depositions in the ng-range and well below. Several types of reference samples were fabricated: multi-elemental layer and extremely low (sub-monolayer) samples for various applications in XRF and total-reflection XRF (TXRF) analysis. Those samples were characterized and compared at three different synchrotron radiation beamlines at the BESSY II electron storage ring employing the reference-free XRF approach based on physically calibrated instrumentation. In addition, the homogeneity of the multi-elemental coatings was checked at the P04 beamline at DESY. The measurements demonstrate the high precision achieved in the manufacturing process as well as the versatility of application fields for the presented reference samples.
\end{abstract}


\end{frontmatter}


\section{Introduction}
\label{S:1}

Improved calibration standards and reference samples of very low mass depositions in the range of ng/mm$^2$ and below are required to enhance the reliability of quantitative X-ray fluorescence based analytical results. Detection limits of commercially available XRF instruments have been steadily decreased while the number of applications in which smallest amounts of materials are to be quantified has similarly increased. Driven by the search for novel material functionalities and improved performance, the variety of investigated nanomaterial combinations with respect to their elemental and structural composition is steadily growing, especially in energy storage applications\cite{J.Azadmanjiri2014} or nanoelectronics\cite{S.W.King2013,Clark2014}. For a reliable quantification with small uncertainties of such low mass depositions, calibration samples need to provide spatially very homogeneous material distributions without any local agglomerations. Unfortunately, the availability of such calibration samples is very limited compared to the quickly growing amount of scientifically and technologically relevant material systems at the nanoscale\cite{G.Roebben2013}.

The ideal reference sample for quantitative XRF analysis in both standard XRF or in total reflection mode XRF (TXRF) geometry incorporates homogeneous distributions of the calibration elements both laterally (in the sample plane) as well as vertically (in the sample depth). The total mass depositions or layer thickness, must be low enough to ensure that both self-absorption and secondary enhancement effects can be neglected. The secondary excitation effects, either by high energy photoelectrons, re-absorbed fluorescence photons or scattered radiation can significantly alter the emitted fluorescence intensities\cite{M.Kolbe2005a} since they directly affect the detection sensitivity. An optimal but unfabricable solution to these issues would be a free standing monolayer of the element of interest.

Earlier work on multi-elemental calibration samples for standard XRF by Pella et al.\cite{P.A.Pella1986} was based on a focused ion beam deposited glass film on a polycarbonate substrate. The glass films were relatively thick (about 0.6 \textmu m) and contained known concentrations of the elements of interest. A correction of the emitted fluorescence radiation for self absorption was still necessary due to the high thickness of the glass layer. In addition, beam damage could limit the life time of the carrier material\cite{G.A.Sleater1987}.

Standard calibration samples for TXRF, e.g. dried droplet samples with known analyte concentration or spin coated calibration samples, can significantly suffer from inhomogeneous elemental distributions, both laterally and vertically\cite{D.Hellin2004,C.Horntrich2012}. This can result in a severe alteration of the emitted fluorescence intensity and thus the calibration reliability\cite{A.Nutsch2009,M.Mueller2012}.

In this work, we present two novel types of samples, which incorporate a reasonable compromise between the mentioned requirements for an ideal calibration sample and the fabricability using deposition techniques common in multilayer production. For standard XRF, mono- or multielemental metal thin films as functional reference materials on thin free standing silicon nitride membranes are presented. For applications in TXRF, very homogeneous layer-like mass depositions on silicon wafer pieces with sub-monolayer depositions as low as 10$^{12}$ atoms/cm$^2$ are introduced. 

The films are produced by physical vapor deposition (PVD) techniques that are widely applied in multilayer production\cite{Rack2010}. These deposition techniques assure very homogeneous layers and a high flexibility regarding the choice of elements and mass densities\cite{Simon2010}. The deposited mass of a certain element and, thus, the signal strength is easily scalable\cite{M.Kraemer2011}.

In comparison with previously mentioned reference samples for TXRF and XRF such as dried droplets (which can become problematic for very low mass depositions due to agglomeration and crystallization in the drying process of the droplet\cite{B.Beckhoff2007a}), bulk pure elements (which may have far too high masses for quantification purposes in the ng range) or borate fluxes (in which the ng range element of interest may be hard to detect), the samples discussed in this publication have a number of advantages for quantitative (\textmu-)XRF and TXRF analyses of very low mass depositions. The very small substrate thickness of the silicon nitride membrane ($\approx$200 nm) and the thin metal deposition with thicknesses ranging from sub-monolayers up to several nanometers provide a quasi absorption free standard for which no matrix correction due to self absorption is necessary. This also results in strongly reduced background contributions providing a good spectral peak-to-background ratio. Experiments in transmission XRF mode are also accessible, allowing for e.g.\ a constant monitoring of the incident beam. Finally, applications in confocal \textmu-XRF setups are possible if the free-standing thin film is located in the common focus of the excitation and detection beam paths. This can be used to align the beam paths or even - with a dedicated layer structure - perform in-depth measurements targeting only one element at a time.

In the standard sample configuration, the fluorescence intensity for all elements is similar, preventing a saturation of the detector by one intense fluorescence line. The metal layers are deposited very homogeneously over the whole sample area.

\begin{figure}[h]
\centering\includegraphics[width=1.\linewidth]{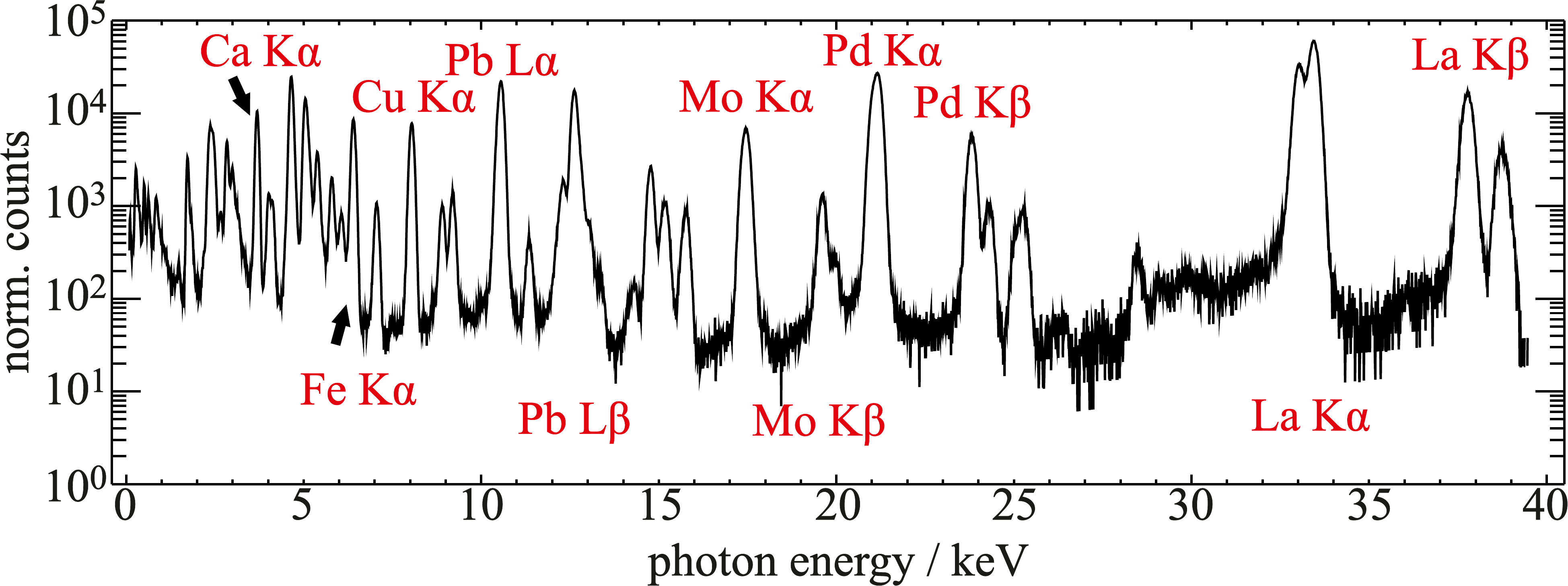}
\caption{\label{fig:SummenS40kV}Sum spectrum of a multi-element reference sample excited with 40 keV photons demonstrating the wide selection of non-overlapping fluorescence lines of similar intensity.}
\end{figure}

The multi-element coating provides a wide selection of non-overlapping intense fluorescence K-lines in the energy range of 3.6 keV (Ca-K$\alpha$) up to 74 keV (Pb-K$\alpha$). Various lines originating from L- and M-shells are also available to extend the range below 3.6 keV.
In Figure \ref{fig:SummenS40kV}
, an exemplary XRF spectrum of such a sample is shown. Fluorescence lines with absorption edges above 40~keV are not observed due to the 40~keV excitation photon energy used for the measurement shown in Figure \ref{fig:SummenS40kV}. The multi-element coating allows for obtaining a calibration curve with many points over a broad energy range with a single measurement and direct quantification of a wide selection of elements or by neighboring elements.

\section{Fabrication and Characterization}

The XRF calibration samples were manufactured as thin multi-elemental depositions on commercially available silicon nitride membranes. The membrane substrates consist of a 10$\times$10~mm$^2$ frame with a usable membrane area of 5$\times$5~mm$^2$ in the center. The metal layers, with thicknesses ranging from sub-monolayers up to several nanometers were deposited directly on the silicon nitride membranes by magnetron sputtering at AXO DRESDEN GmbH.  Two membranes were coated with seven different elements each. Ca, Fe, Cu, Mo, Pd, La and Pb were sputtered on both samples with the second sample having a ten times lower mass deposition than the first one. 

The mass deposition on the samples was tuned by adjusting the power at which the magnetrons are run as well as by the movement of the sample holder over the magnetron targets. Further, tests were carried out to determine the ideal order and mass of the materials in the stack taking into account possible chemical reactions between certain elements. The interface roughness - usually a critical point in multilayer production - did not play a role here as only the total mass in the sample system was important for the X-ray fluorescence signal. However, the lateral homogeneity over the sample area of interest has to be very good, which was provided by the optimized instrumental settings.

Additionally, various mono-elemental coatings with thicknesses of several ten nanometers have been fabricated and were used for experimental determinations of atomic fundamental parameters \cite{M.Kolbe2012,P.Hoenicke2014,M.Kolbe2015}. Here, the thickness homogeneity as well as a low surface roughness were very important. These parameters were ensured by measuring test samples coated in the same deposition run.

Silicon wafers, rather than silicon nitride membranes, were chosen as substrates for the TXRF reference samples for two main reasons. First, the grazing incidence angle in TXRF is very small, leading to a beam footprint of several centimeters, thus being larger than available silicon nitrate membranes. Second, ultrapure industrial silicon wafers are commercially available with extremely low contaminations. This is necessary if mass depositions down to the picogram range are to be measured while ensuring that no unintentional contaminations at a relevant level are present. Nickel was selected as coating element because it hardly occurs as contamination on ultrapure industrial silicon wafers and, thus, only well-defined material amounts are present on the sample. Being a magnetic material, it is difficult to sputter nickel with magnetrons. Thus, Dual Ion Beam Deposition (DIBD) was selected as a fabrication technique. In DIBD an ion beam sputters material from a target onto a sample carrier. A second ion beam in the same machine can be used for polishing/cleaning of the substrate as well as adding additional energy to the particles in the growing films. With typical energies between 10 eV and 100 eV (in comparison to magnetron sputtered atoms of around 1 eV to 10 eV), this method reduces island formation and provides very low surface roughness. Another advantage of DIBD is that critical process parameters can be tuned continuously and thus scaled down smoothly and reproducibly. Several runs of test samples were fabricated and measured to optimize the machine parameters and to meet the  mass depositions intended. Standard Si wafer pieces were coated with Ni at nominal mass depositions between $9\cdot10^{11}$ atoms/cm$^2$ and $9\cdot10^{15}$ atoms/cm$^2$. Two independent sets of samples were produced (cf. Table \ref{tab:TXRFsamples}). 

\begin{table}[htbp]
  \centering
  \caption{List of the TXRF calibration samples fabricated in this study and nominal layer thicknesses as well as nominal atomic mass depositions. Each set contains a pair of identical nominal layer thickness to study the run-to-run stability of the coating process.}
    
\begin{tabular}{|c|c|c|c|}
\hline
    set & sample  & d$_{Ni}$ / nm & mass dep. / at. cm$^{-2}$ \bigstrut\\
\hline
    \multirow{5}[10]{*}{A} & TX-K09-01 & 1     & 9.1E+15 \bigstrut\\
\cline{2-4}                & TX-K09-02 & 0.1   & 9.1E+14 \bigstrut\\
\cline{2-4}                & TX-K09-03 & 0.01  & 9.1E+13 \bigstrut\\
\cline{2-4}                & TX-K09-04 & 0.01  & 9.1E+13 \bigstrut\\
\cline{2-4}                & TX-K09-05 & 0.005 & 4.6E+13 \bigstrut\\
    \hline
    \multirow{6}[12]{*}{B} & TX-K47-09 & 0.01  & 9.1E+13 \bigstrut\\
\cline{2-4}                & TX-K47-10 & 0.005 & 4.6E+13 \bigstrut\\
\cline{2-4}                & TX-K47-11 & 0.001 & 9.1E+12 \bigstrut\\
\cline{2-4}                & TX-K47-12 & 0.001 & 9.1E+12 \bigstrut\\
\cline{2-4}                & TX-K47-13 & 0.0005 & 4.6E+12 \bigstrut\\
\cline{2-4}                & TX-K47-14 & 0.0001 & 9.1E+11 \bigstrut\\
    \hline
    \end{tabular}%
    
  \label{tab:TXRFsamples}%
\end{table}%

Both the multi-elemental as well as the TXRF reference samples were characterized using transmission and fluorescence measurements at PTB employing radiometrically calibrated instrumentation and the atomic fundamental parameter based reference-free quantification approach\cite{Beckhoff2008}. Thus, the mass depositions of the coated materials can be determined without the need for any calibration sample. 

The reference-free XRF experiments were carried out at three different beamlines at the electron storage ring BESSY II. In addition, homogeneity experiments were performed at the P04 beamline at DESY. In total, an incident photon energy range of 78~eV up to 80~keV is covered by the used beamlines. The plane grating monochromator (PGM) beamline\cite{F.Senf1998} for undulator radiation provides soft X-ray radiation of high spectral purity in the photon energy range of 78~eV to 1860~eV. Hard X-ray radiation between 1.75~keV and 10.5~keV is available at the four-crystal monochromator (FCM) beamline for bending magnet radiation\cite{Krumrey1998}. Both beamlines are located in the PTB laboratory at BESSY II\cite{B.Beckhoff2009c}. Additionally, experiments employing radiation with photon energies above 10 keV were carried out at a 7-T wavelength shifter (WLS) beamline\cite{W.Goerner2001}. The variable polarization XUV beamline P04 at DESY\cite{VIEFHAUS2013151} with an APPLE-2 undulator covers the tender energy range from 200~eV to 3~keV with very high brilliance and energy resolution offered by varied line space gratings.

\begin{figure}[h]
\centering\includegraphics[width=0.5\linewidth]{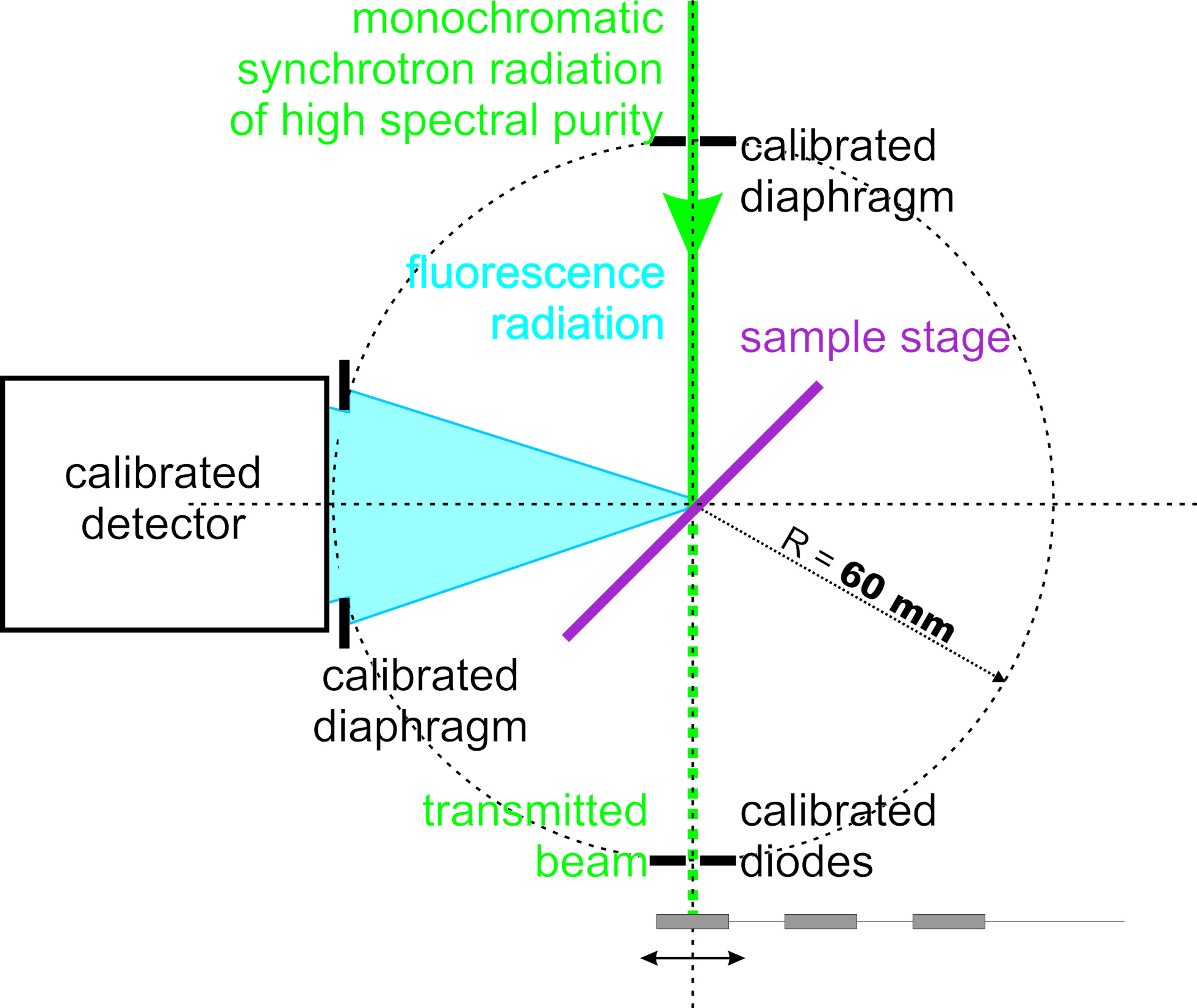}
\caption{\label{fig:XRF_schematic}Schematic view of the experimental set up at the PTB beamlines for the characterization of the XRF reference samples.}
\label{XRF_schematisch_eng}
\end{figure}

The experiments at all three beamlines at BESSY II were carried out in ultra high vacuum chambers, equipped with calibrated photo diodes and an energy-dispersive silicon drift detector (SDD) with known response functions and detection efficiency\cite{F.Scholze2009}. Each sample can be placed into the center of the respective chamber by means of an x-y scanning stage. The incident angle $\Psi_{in}$ between the surface of the sample and the incoming beam was set to 45$^\circ$ for the experiments using the XRF calibration samples. This setup is described in detail in \cite{M.Kolbe2005a} and a schematic view is shown in Figure \ref{fig:XRF_schematic}. 
For the TXRF calibration samples, a different experimental setup allowing for experiments in total reflection and in grazing incidence (GI) mode was used. This instrument\cite{J.Lubeck2013} allows for the necessary realization of very small angles of incidence between the sample surface and the exciting beam. In addition, the accurate variation of the incident angle in a broad angular range can be ensured. 
The experiments at the P04 beamline at DESY were carried out in a high vacuum chamber, equipped with a zone plate as focussing optic, a scanning device with sub nanometer resolution and a specially designed 4-channel SDD with a solid angle of detection of up to 1.2~sr\cite{andrianov2017scanning}. The setup is a typical setup for scanning X-ray microscopy with an incident angle $\Psi_{in}$ set to 90$^\circ$ and the SDD detector aligned in backscatter geometry.

\section{Results and discussion}
\subsection{XRF calibration samples}
The fabricated samples were analyzed to characterize both the lateral homogeneity of the deposition process across the membrane area and the scalability as well as the absolute mass deposition coated on the substrate. Those were verified by comparing the quantitative results of each element determined by reference-free XRF analysis for the two samples with a nominal mass deposition ratio of 10. The mass depositions were derived from XRF spectra, recorded with two different incident photon energies of 5.75~keV and 13.5~keV, respectively, to provide optimized excitation conditions for all elements. The recorded fluorescence spectra were deconvoluted using relevant bremsstrahlung contributions and the detector response functions of the fluorescence lines of interest. The reference-free quantification of the elemental mass depositions was performed using the Sherman equation\cite{Sherman1955}. All relevant instrumental parameters, e.g.\ the incident photon flux or the solid angle of detection are known due to the use of our physically traceable calibrated instrumentation\cite{M.Kolbe2005a,Beckhoff2008}. The atomic fundamental parameters, e.g.\ the fluorescence yields, the Coster-Kronig factors or the photoionization cross sections have been experimentally determined using the monoelemental coatings\cite{M.Kolbe2012,P.Hoenicke2014,M.Kolbe2015,Menesguen2017} or were taken from databases\cite{T.Schoonjans2011,H.Ebel2003,Zschornack}. The results of the reference-free quantification are shown in Table \ref{tab:quant}.

\begin{table}[htbp]
  \centering
  \caption{Comparison of the determined mass depositions for the two multi elemental samples of 7 different coating elements using reference-free XRF. Two different incident photon energies (E$_0$) were chosen for optimized excitation conditions.}
    \begin{tabular}{|c|c|c|ccc|ccc|ccc|}
    \hline
    \multicolumn{1}{|p{2.5em}|}{} & \multicolumn{1}{c|}{\textbf{Lineset}} & \textbf{E$_0$} & \multicolumn{3}{c|}{\textbf{Multi 0.1x}} & \multicolumn{3}{c|}{\textbf{Multi 1x}} & \multicolumn{3}{c|}{\textbf{Ratio}} \bigstrut\\
    \hline
    \multicolumn{1}{|c|}{\textbf{La}} & \multicolumn{1}{c|}{\textbf{L3}} & \textbf{5.75} & 12.9 & $\pm$     & 1.1  & 122.7 & $\pm$     & 10.2  & \multicolumn{1}{c}{9.6} & \multicolumn{1}{c}{$\pm$} & 1.2 \bigstrut\\
    \hline
    \multicolumn{1}{|c|}{\textbf{Pd}} & \multicolumn{1}{c|}{\textbf{L3}} & \textbf{5.75} & 4.3  & $\pm$     & 0.4  & 36.6 & $\pm$     & 2.7  & \multicolumn{1}{c}{8.6} & \multicolumn{1}{c}{$\pm$} & 1.0 \bigstrut\\
    \hline
    \multicolumn{1}{|c|}{\textbf{Mo}} & \multicolumn{1}{c|}{\textbf{L3}} & \textbf{5.75} & 0.56  & $\pm$     & 0.07  & 5.0  & $\pm$     & 0.4  & \multicolumn{1}{c}{9.1} & \multicolumn{1}{c}{$\pm$} & 1.3 \bigstrut\\
    \hline
    \multicolumn{1}{|c|}{\textbf{Fe}} & \multicolumn{1}{c|}{\textbf{K}} & \textbf{13.5} & 4.53  & $\pm$     & 0.31  & 40.3 & $\pm$     & 2.8  & \multicolumn{1}{c}{8.9} & \multicolumn{1}{c}{$\pm$} & 0.9 \bigstrut\\
    \hline
    \multicolumn{1}{|c|}{\textbf{Cu}} & \multicolumn{1}{c|}{\textbf{K}} & \textbf{13.5} & 2.4  & $\pm$     & 0.2  & 22.6  & $\pm$     & 1.7  & \multicolumn{1}{c}{9.3} & \multicolumn{1}{c}{$\pm$} & 1.0 \bigstrut\\
    \hline
    \multicolumn{1}{|c|}{\textbf{Ca}} & \multicolumn{1}{c|}{\textbf{K}} & \textbf{5.75} & 18.7 & $\pm$     & 1.6  & 171.9 & $\pm$     & 14.2 & \multicolumn{1}{c}{9.2} & \multicolumn{1}{c}{$\pm$} & 1.1 \bigstrut\\
    \hline
    \multicolumn{1}{|c|}{\textbf{Pb}} & \multicolumn{1}{c|}{\textbf{L3}} & \textbf{13.5} & \multicolumn{1}{c}{7.1} & $\pm$     & 0.5  & 74.3 & $\pm$     & 5.2  & \multicolumn{1}{c}{10.6} & \multicolumn{1}{c}{$\pm$} & 1.1 \bigstrut\\
    \hline
          &       & \textbf{keV} & \multicolumn{3}{c|}{\textbf{ ng mm$^{-2}$}} & \multicolumn{3}{c|}{\textbf{ng mm$^{-2}$}} & \multicolumn{3}{c|}{} \bigstrut\\
    \hline
    \end{tabular}%
  \label{tab:quant}%
\end{table}%

The determined mass ratios for each element are well in line with the nominal target value of 10 with respect to the experimental uncertainties. The experimental uncertainties are in the order of 8\%. Main contributors to the uncertainty budget are the uncertainties of the relevant fundamental parameters employed. 

\begin{figure}[h]
\centering\includegraphics[width=\linewidth]{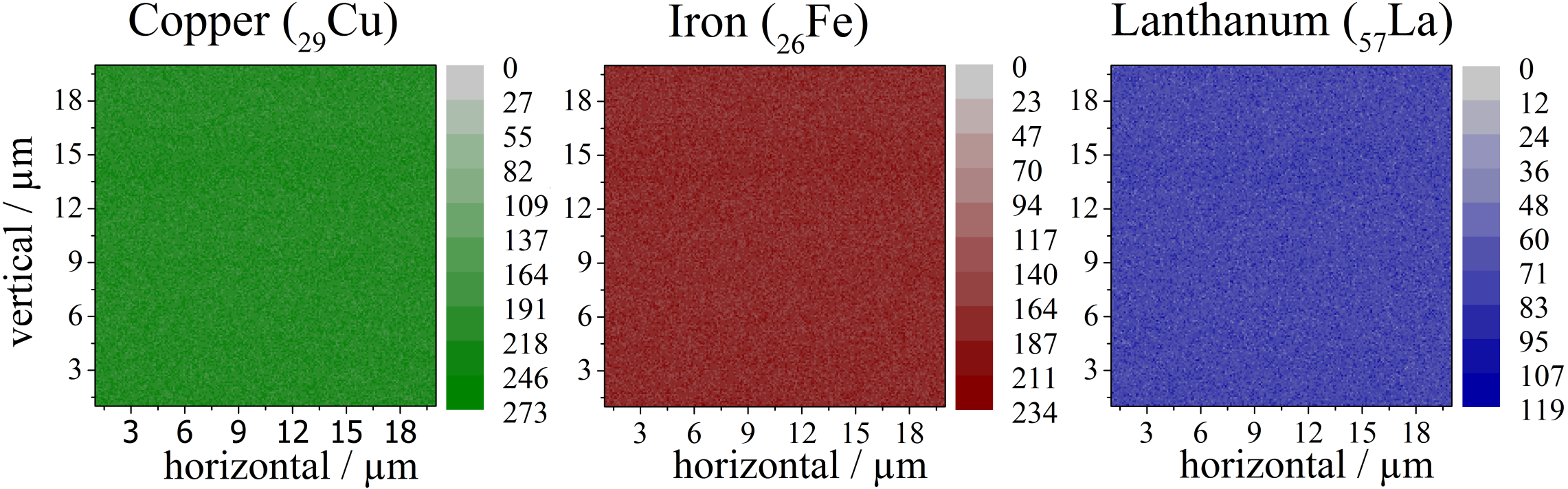}
\caption{Mapping results derived from the single spectra by region of interest integration for Cu-L (green), Fe-L (red) and La-M (blue). The pixel size is approximately 100$\times$100~nm$^2$. See text for further details.}
\label{fig:XRF_Scan_P04}
\end{figure}

The lateral homogeneity of the samples was determined by scanning across the free-standing membrane area with a small incident beam and recording the excited fluorescence radiation of the elements of interest. The combination of the high brilliance of the P04 beamline at DESY and the large solid angle of the specially designed 4-channel SDD detector was used to map several areas of the coated membrane with a spot size of approximately 100$\times$100~nm$^2$. Acquisition times varied between 5~ms and 20~ms per measurement and a time based continuous scanning mode was used. An excitation photon energy of 1~keV was used in order to excite fluorescence radiation from the M- and L-shells of several of the coating elements. To evaluate the lateral homogeneity, regions of interest (ROI) around the characteristic fluorescence lines from the recorded spectra for Fe-L, La-M and Cu-L were evaluated. In Figure \ref{fig:XRF_Scan_P04}, this data is shown for an acquisition time of 20 ms  and a step size of 100 nm.

Within the counting statistics of the derived ROI intensities the coating for the various metals is homogeneous. This is confirmed by the comparison of the standard deviation $\sigma$ of all measurement positions with the counting statistics of the mean value $\sqrt[]{\bar{X}_{ROI}}$. The difference of $\sigma$ and $\sqrt[]{\bar{X}_{ROI}}$ is less than 1.5\% for all elements in the inspected areas.

\subsection{TXRF calibration samples}

The TXRF calibration samples were characterized regarding their homogeneity in both the lateral and vertical directions. In the ideal case, after the deposition Ni atoms are located on the Si wafer surface without any agglomeration or cluster formation. This can be monitored using reference-free grazing incidence XRF (GIXRF)\cite{M.Mueller2014} varying the incident angle around the critical angle for total external reflection. On flat samples, the interference between the incoming and the reflected beam results in the so called X-ray standing wave (XSW) field. The intensity distribution inside the XSW strongly depends on the incident angle. Thus, angular variation of the emitted fluorescence lines reveals information about the vertical distribution of the element of interest\cite{M.Kraemer2006,Kraemer2007,P.Hoenicke2009}. 

\begin{figure}[h]
\centering\includegraphics[width=1.\linewidth]{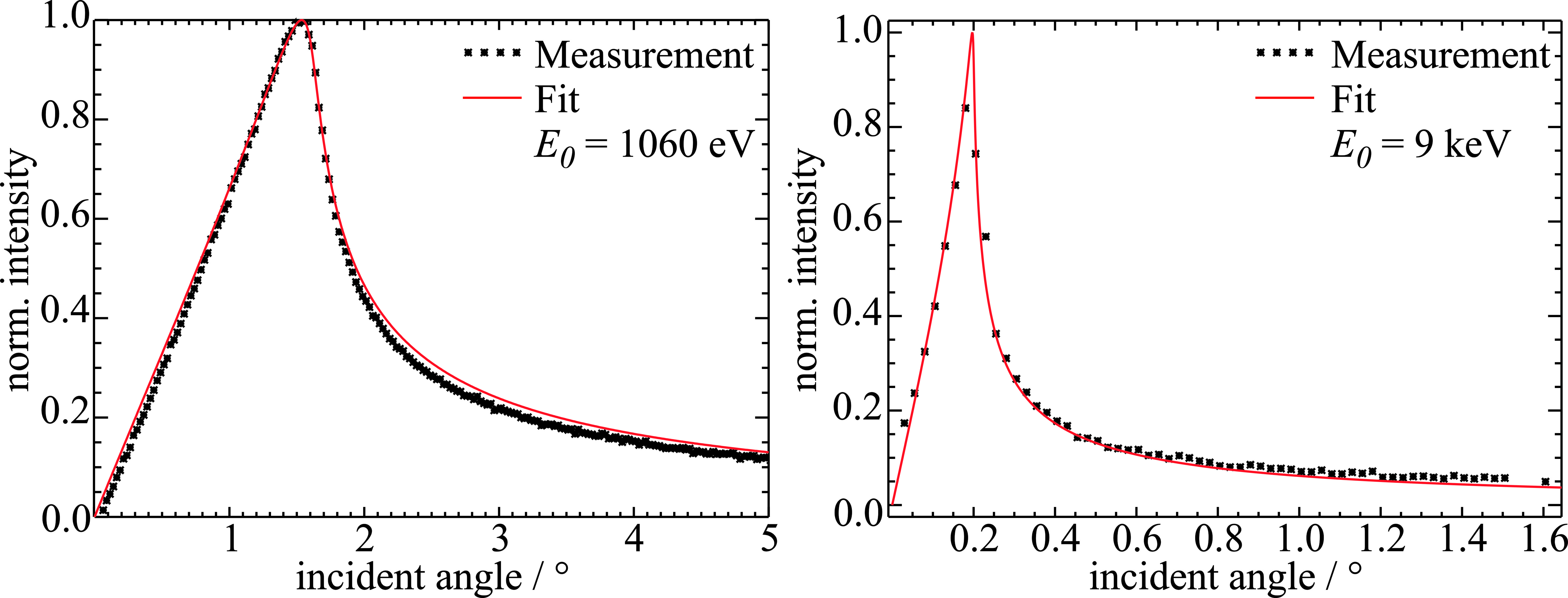}
\caption{Comparison of GIXRF measurements on two TXRF calibration samples TX-K09-03 (left hand side, measured using Ni-L$\alpha$) and TX-K47-11 (right hand side, measured using Ni-K$\alpha$) with a targeted Ni thickness of $0.01$ nm and $0.001$ nm, respectively. The theoretical fits assuming a thin layer-like Ni distribution on the sample surface agree very well with the measured curves for both samples.}
\label{fig:Vergleich_TXRF_Fit}
\end{figure}

In Figure \ref{fig:Vergleich_TXRF_Fit}, GIXRF results of TXRF calibration samples with nominal Ni mass depositions of $9\cdot10^{13}$ atoms/cm$^2$ (sample TX-K09-03, left hand side) and $9\cdot10^{12}$ atoms/cm$^2$  (sample TX-K09-11, right hand side) are shown. The measurement of sample TX-K09-03 was carried out at the PGM beamline exciting Ni-L fluorescence lines with incident photons of 1.06~keV. Sample TX-K09-11 was excited with 9~keV photons at the FCM beamline exciting also the K-fluorescence lines of Ni. The fitted angular profiles, simulated with a thin Ni layer on the surface of the silicon wafer, show good agreement with the measurements. Agglomeration, clustering or diffusion would result in deviations with respect to the fitted curve at either low or high incident angles.

In order to determine the lateral homogeneity of the Ni deposition, the angular scans were repeated at different positions across the sample surface for two samples. The resulting angular fluorescence profiles reveal that, first of all, no significant variations in the position and shape of the peak are observed. Secondly, the fluorescence intensity at high incident angles, which is proportional to the local mass deposition of Ni, is within a range of $\pm$2\% for all measurement points on one sample. This indicates that the deposited nickel is distributed homogeneously over the sample surface without forming any islands or clusters.

\begin{figure}[h]
\centering\includegraphics[width=0.8\linewidth]{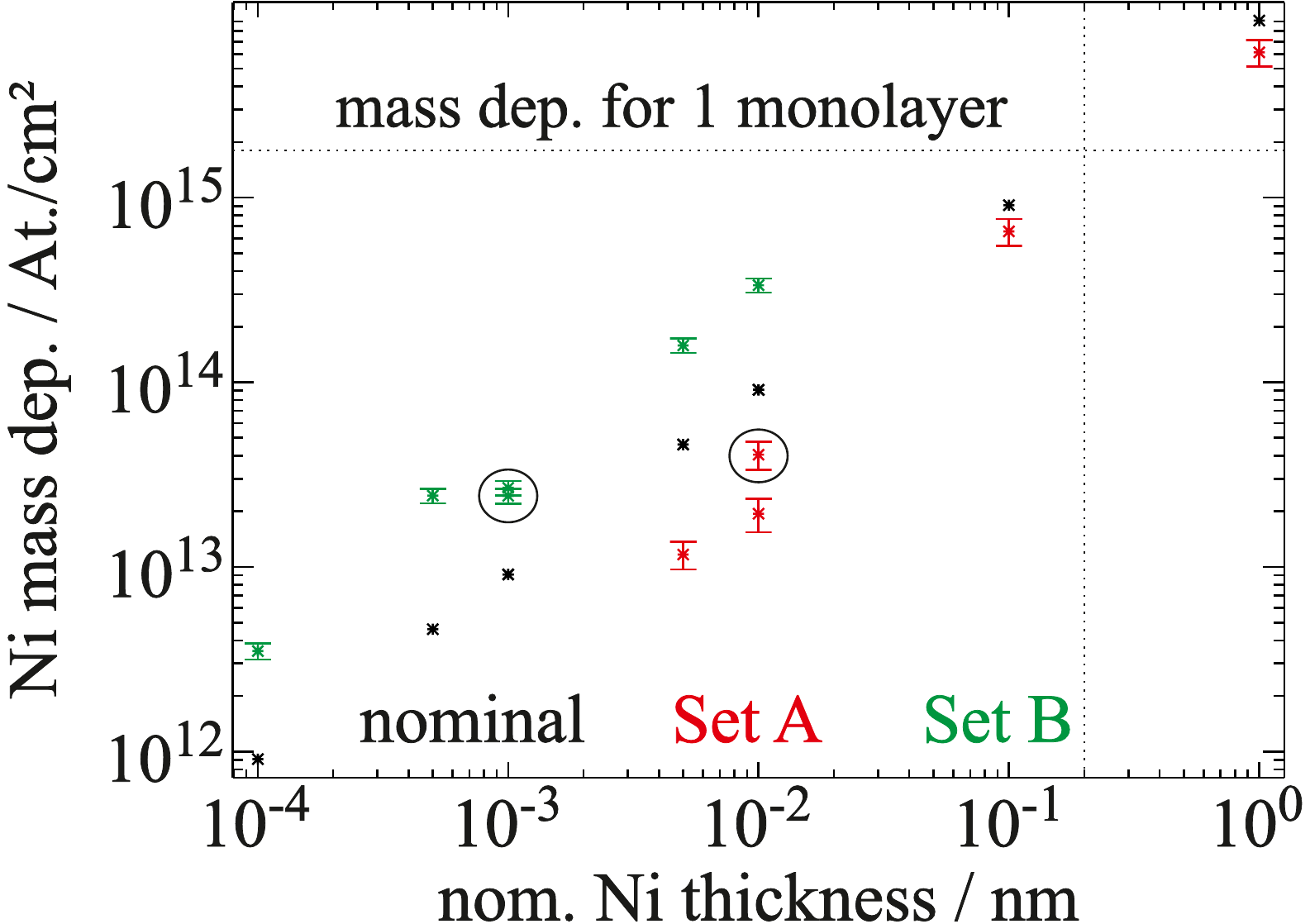}
\caption{Comparison of the measured Ni mass depositions in comparison to the nominal values for both sample sets investigated. The two circled results correspond to the samples for which the fits in fig. \ref{fig:Vergleich_TXRF_Fit} are shown.}
\label{fig:Vergleich_TXRF_Quant}
\end{figure}

Linearity and reproducibility of the deposition process were evaluated by quantification of the total Ni mass depositions from the reference-free GIXRF\cite{M.Mueller2014} experiments for both sample sets (details will be described later). The nominal mass depositions were calculated using the nominal layer thickness and tabulated mass densities $\rho$. A comparison of the quantified Ni mass depositions with the respective nominal values is shown in Figure \ref{fig:Vergleich_TXRF_Quant} where the results for all samples are plotted in comparison to the nominal values. The corresponding quantification results are shown in table \ref{tab:TXRFresults}. It should be noted, that the uncertainties of the quantification are higher for sample set A, which was measured at the PGM beamline exciting the Ni-L fluorescence lines. The respective fundamental parameters\cite{Guerra2018}, which are the main contributors to the quantification uncertainty, have higher uncertainties as compared to the K-shell and, in addition, also Coster-Kronig transitions must be taken into account.

\begin{table}[htbp]
  \centering
  \caption{Comparison of the quantification results on the TXRF calibration samples to the nominal values.}
    
\begin{tabular}{|c|c|c|c|}
\hline
    \multirow{2}[1]{*}{set} & \multirow{2}[1]{*}{sample} & nominal & quantified \bigstrut\\
     &   & mass dep. / at. cm$^{-2}$ & mass dep. / at. cm$^{-2}$ \bigstrut\\
\hline
    \multirow{5}[10]{*}{A} & TX-K09-01 & 9.1E+15 & 6.1(1.0)E+15 \bigstrut\\
\cline{2-4}                & TX-K09-02 & 9.1E+14 & 6.6(1.1)E+14 \bigstrut\\
\cline{2-4}                & TX-K09-03 & 9.1E+13 & 4.1(0.7)E+13 \bigstrut\\
\cline{2-4}                & TX-K09-04 & 9.1E+13 & 1.9(0.4)E+13 \bigstrut\\
\cline{2-4}                & TX-K09-05 & 4.6E+13 & 1.2(0.2)E+13 \bigstrut\\
    \hline
    \multirow{6}[12]{*}{B} & TX-K47-09 & 9.1E+13 & 3.4(0.3)E+14 \bigstrut\\
\cline{2-4}                & TX-K47-10 & 4.6E+13 & 1.6(0.14)E+14 \bigstrut\\
\cline{2-4}                & TX-K47-11 & 9.1E+12 & 2.4(0.22)E+13 \bigstrut\\
\cline{2-4}                & TX-K47-12 & 9.1E+12 & 2.7(0.24)E+13 \bigstrut\\
\cline{2-4}                & TX-K47-13 & 4.6E+12 & 2.4(0.22)E+13 \bigstrut\\
\cline{2-4}                & TX-K47-14 & 9.1E+11 & 3.5(0.35)E+12 \bigstrut\\
    \hline
    \end{tabular}%
    
  \label{tab:TXRFresults}%
\end{table}%

Both sets show a linear scalability down to very low mass depositions. Due to the very small thickness of the deposited “layer”, direct measurement during the deposition process was not possible. Thus, a deviation (offset) from the nominal values occurred with the coated amounts of set A being lower than the nominal target values. The Ni amount was determined by synchrotron radiation based reference-free XRF after completion of set A. Based on the reference-free XRF results, the coating process was adjusted to provide (a) a better run-to-run stability, (b) a better fit to the nominal target value and (c) a scalability down to even lower mass depositions as the method used for set A was limited to a minimum of some picometers. 

Some improvements made are the fabrication of a better and more homogeneous Ni target to provide more flexibility in the scaling down process of the target masses. Further, deposition slits as well es substrate movements were tuned in a new mechanical set-up to easily target the range of 1 to 10 picometers and have degrees of freedom to move further down by a few orders of magnitude.

The second series of Ni layers (set B) shows a good linearity and a much better run-to-run stability than set A. The coated values, however, are a bit higher than the target values. This is caused by the above-mentioned changes in the fabrication and the lack of measurement capabilities during the coating process (which could help reduce the effect). Even  though the measured values do not agree exactly with the nominal (target) values, this is not necessary for the purpose of these reference samples. With the reproducibility and precision of the coating method a large number of substrates can be coated in one or a few subsequent runs of which only one test sample has to be characterized by synchrotron-based reference-free XRF measurements. These samples can then be used as reference samples with known mass deposition.

\section{Conclusions}
In the present work, deposition techniques widely used for the production of multilayer mirrors have been successfully applied to the production of XRF calibration samples. The two presented types of calibration samples for quantitative XRF investigations are both very close to an ideal calibration sample. The low mass depositions as well as the thin membrane backing of the XRF samples allow to neglect self absorption and secondary excitation effects. Furthermore, the spectral background is very low and it was shown that the lateral homogeneity of the deposited metals is very high. This enables the routine use of these multi-elemental reference samples for the full bandwidth of XRF applications, i.e. nano-, micro-XRF and standard XRF.

Further improvement of the reproducibility as well as a continued down scaling of the deposited Ni areal mass for the TXRF samples is planned. However, the current status is already well suited for an application as a TXRF calibration sample because these samples allow for a calibration of a TXRF instrument without the potential introduction of errors associated with an unevenly dried or crystallized droplet sample. Due to the good scalability of the deposition, several orders of magnitude for the targeted mass deposition can be covered. 

\section{Acknowledgements}
The authors would like to thank J. Weser, M. Gerlach, M. Kolbe (all PTB) and R. Britzke (Bundesanstalt f\"ur Materialforschung und -pr\"ufung, BAM) for their support during the experiments. The financial support by the MNPQ-Transfer program of the German Federal Ministry of Economics and Technology is also gratefully acknowledged. Parts of this research were performed within the EMPIR project 3D-Stack. The financial support of the EMPIR program is gratefully acknowledged. It is jointly funded by the European Metrology Program for Innovation and Research (EMPIR) and participating countries within the European Association of National Metrology Institutes (EURAMET) and the European Union.

\section{References}


\bibliographystyle{model1-num-names}
\bibliography{sample.bib}







\end{document}